\documentclass[useAMS,usenatbib]{mn2e}
\usepackage{graphicx}
\newcommand{\etal}{{\it et al.} }
\newcommand{\asca}{{\it ASCA} }

\newcommand{\xmm}{{\it XMM-Newton} }
\newcommand{\xmmp}{{\it XMM-Newton}}
\newcommand{\chandra}{{\it Chandra} }

\newcommand{\rxte}{{\it RXTE} }
\newcommand{\hetg}{{\it HETGS} }
\newcommand{\letg}{{\it LETGS} }

\newcommand{\bsax}{{\it BeppoSAX} }

\newcommand{\silya}{Si~{\sc xiv}~Ly$\alpha$ }
\newcommand{\nelya}{Ne~{\sc x}~Ly$\alpha$ }
\newcommand{\nlya}{N~{\sc vii}~Ly$\alpha$ }
 
\newcommand{\oxla}{O~{\sc viii}~Ly$\alpha$ }
\newcommand{\mglya}{Mg~{\sc xii}~Ly$\alpha$ }

\newcommand{\oxyseven}{O~{\sc vii} }
\newcommand{\oxyeight}{O~{\sc viii} }
\newcommand{\oxysevenf}{[O~{\sc vii}] }
\newcommand{\oxyseveni}{O~{\sc vii} (i) }
\newcommand{\oxysevenr}{O~{\sc vii} (r) }
\newcommand{\nenine}{Ne~{\sc ix} }
\newcommand{\neninef}{[Ne~{\sc ix}] }
\newcommand{\neniner}{Ne~{\sc ix} (r) }

\newcommand{\mgelevenr}{Mg~{\sc xi} (r) }

\newcommand{\sithirteenr}{Si~{\sc xiii} (r) }

\newcommand{\thr}{3C~120 }

\newcommand{\mcg}{MCG~$-$6$-$30$-$15 }

\title[A Soft X-ray study of type I AGN observed with Chandra]{A SOFT X-RAY STUDY OF TYPE~I AGN OBSERVED WITH \chandra \hetg}

\author[B. McKernan, T. Yaqoob \& C. S. Reynolds]{B. McKernan$^{1,2}$\thanks{E-mail:mckernan@astro.umd.edu (BMcK)}, T. Yaqoob$^{3,4}$ and C. S. Reynolds$^{5}$\\
$^{1}$Department of Science, Borough of Manhattan Community College, City University of New York, New York, NY 10007\\
$^{2}$Hayden Associate, Hayden Planetarium, Dept. of Astrophysics, American Museum of Natural History, New York, NY 10024\\
$^{3}$Department of Physics and Astronomy,
                        Johns Hopkins University, Baltimore, MD 21218\\
$^{4}$Laboratory for High Energy Astrophysics,
                NASA/Goddard Space Flight Center, Greenbelt, MD 20771\\
$^{5}$Department of Astronomy, University of Maryland, 
College Park, MD 20742}
\begin{document}

\date{Accepted. Received; in original form}

\pagerange{\pageref{firstpage}--\pageref{lastpage}} \pubyear{2007}

\maketitle

\label{firstpage}

\begin{abstract}
We present the results of a uniform analysis of the soft X-ray spectra of 
fifteen type I AGN observed with the high resolution X-ray gratings on 
board \emph{Chandra}. We found that ten of the fifteen AGN exhibit 
signatures of an intrinsic ionized absorber. The absorbers are 
photoionized and outflowing, with velocities in the range 
$\sim 10^{1}-10^{3}$ km $\rm{s}^{-1}$. The column density of the warm 
absorbing 
gas is $\sim 10^{20-23} \rm{cm}^{-2}$. Nine of the ten AGN exhibiting 
warm absorption are best--fit by multiple 
ionization components and three of the ten AGN \emph{require} multiple 
kinematic components. The warm absorbing gas in our AGN sample has a 
wide range of ionization parameter, spanning roughly four orders of magnitude 
($\xi \sim 10^{0-4}$ ergs cm $\rm{s}^{-1}$) in total, and often spanning 
three orders of magnitude in the same gas. Warm absorber components with 
ionization parameter $\xi<10$ generate an unresolved transition 
array due to Fe in seven of the ten AGN exhibiting warm absorption. These 
low ionization state absorbers may also carry away the largest mass outflows
 from the AGN. The mass outflow rate depends critically
on the volume filling factor of the gas, which cannot yet be directly 
measured. However, upper limits on the mass outflow rates
for filling factors of unity 
can be much greater than the 
expected accretion rate onto the central supermassive 
black hole and filling factors as small as 1\% can give
outflow rates comparable to the accretion rate.
There appears to be a gap in the outflow velocities
in our sample between $\sim 300-500$ km $\rm{s}^{-1}$, the origin of
which is not clear. The outflow components with velocities below this
gap tend to be associated with lower column densities than those with
with velocities above the gap. 
 
\end{abstract}


\section{Introduction}
\label{sec:intro}
    X-ray emission from active galactic nuclei (AGN) is believed to be
 powered by an accretion flow onto a supermassive black hole (SBH). In
the unified AGN paradigm, type~I AGN have accretion disks at
  a small angle of inclination to the observers' line--of--sight. Hard
 X-ray ($>$2 keV) spectra from type~I AGN typically reveal a continuum
 that is well described by a simple cut--off power--law model, with a 
fluorescent
           Fe~K line complex (and sometimes a reflection continuum)
 superimposed (see eg. \citet{b36} and references therein).  
The soft X-ray ($<$2 keV) spectra of type~I AGN are typically complex 
and can yield a great deal of information about the distribution and 
state of matter in the AGN central engine.

\begin{table*}
 \begin{minipage}{155mm}
   \caption{The \chandra \hetg sample of type~I AGN \label{tab:sample}. 
AGN are listed in order of increasing RA. Columns 2,3 and 4 give 
the co-ordinates and the redshift of the 
source (from  NED). Redshift was deduced from observations of the 
21cm H~{\sc{i}} line where possible, since optical estimates of $z$ 
may be confused by AGN outflow. Column 7 is the mean count rate for the 
combined $\pm1$ order spectra of the Medium Energy Grating (MEG). $^{a}$ from \citet{b10}, except for Mkn~509 \citet{b27}.  The Galactic column 
density towards F9, NGC~7314, NGC~3516, NGC~3783, NGC~5548, Mkn 766, NGC~3227 
and  Akn 564 was obtained from \citet{b35}. $^{b}$ NGC~3516: Dates of three 
`snapshots' are given; NGC~3783: data were combined from five observations 
over a period of $\sim 124$ days; MCG$-$6-30-15: this observation was made 
over a period of $\sim 138$ days in three parts and only the combined data 
are analyzed here. $^{c}$ Total good integration time of spectrum. 
$^{d}$ \thr is also classified as a broad--line radio galaxy (NED). $^{e}$ To 
gauge changes in AGN spectral continuua, particularly in the vicinity of the 
Oxygen absorption edges, the \xmm EPIC PN spectrum of this AGN was studied. 
$^{f}$ we summed the first two observations of NGC 3516
into a single `low state' spectrum and compared this with the `high state' 
spectrum in the third observation.}
   \begin{tabular}{@{}lrrrcllr@{}} 
   \hline
Source & RA & Decl. & Redshift & Galactic $N_{H}$ & Observation & MEG count &
Exposure$^{c}$ \\
 & (J2000.0) & (J2000.0) & (z) & 
($10^{20} \rm{cm}^{-2}$) $^{a}$ &Start $^{b}$& rate(ct/s) & (ks)\\
\hline
  Fairall 9 & 01 23 45.7 & -58 48 21 & 0.04600 & 3.0& 2001 Sep 11 & $0.486\pm 0.02$ &80
\\  
3C 120$^{d}$ & 04 33 11.0 & 05 21 15 & 0.03301 & 12.30 & 2001 Dec
   21 & $0.814 \pm 0.004$ &58 \\ 
NGC~3227 & 10 23 30.6 & +19 51 54 & 0.00386 & 2.15 & 1999 Dec 30
& $0.130\pm 0.003$ &47 \\
NGC~3516 & 11 06 47.5 &+72 34 07 & 0.00884 & 3.05 & 2001 April 9 & &36 \\
         &            & & & & 2001 April 10 &$0.132\pm 0.002$$^{f}$ &75 \\
         &            & & & & 2001 Nov 11&$0.230\pm 0.003$ &89 \\
NGC 3783 &11 39 01.7 & -37 44 18 & 0.00973 &8.50 & 2001
  Feb 24 & $0.668\pm 0.002$ &850 \\ 
NGC 4051 & 12 03 09.5 & 44 31 52 & 0.00242 & 1.31 &
2000 Mar 24 & $0.396 \pm 0.002$ &80 \\ 
Mkn 766 & 12 18 26.5 & 29 48 46 & 0.01293 & 1.80&
2001 May 7 & $0.572 \pm 0.002$ &90 \\ 
NGC 4593$^{e}$ & 12 39 39.3 & -05 20 39 & 0.00831 & 1.97
 & 2001 Jun 29 & $0.948 \pm 0.004$ &79 \\ 
MCG-6-30-15$^{e}$ & 13 35 53.3 & -34 17 48 & 0.00775
       &4.06& 2000 Apr 5 & $0.594 \pm 0.002$ &126 \\ 
IC 4329A$^{e}$ & 13 49 19.2 & -30 18 34 &
0.01605 & 4.55 & 2001 Aug 26 & $2.244 \pm 0.006$ &60 \\ 
Mkn~279 & 13 53 03.5 & +69 18 30 & 0.03045 & 1.64 & 2002 May 18 & $0.241\pm 
0.002$ &116 \\
NGC 5548$^{e}$ & 14 17 59.5 & 25 08 12
  & 0.01717 & 1.70 & 2000 Feb 5 & $0.403 \pm 0.002$ &82 \\ 
Mkn 509$^{e}$ & 20 44 09.6 & -10 43
  23 & 0.03440 & 4.44 & 2001 Apr 13 & $0.963 \pm 0.004$ &60 \\ 
NGC~7314 & 22 35 46.2 & -26 03 01 & 0.00474 & 1.46 & 2002 Jul 19 & $0.353\pm 
0.002$&97 \\
Akn 564 & 22 42 39.3 & 29
                         43 31 & 0.02467 & 6.40 & 2000 Jun 17 & $1.302 \pm 
0.007$ &50 \\
\hline
\end{tabular}
\end{minipage}
\end{table*}

Spectral complexity beyond a simple power law model in the 
soft X--ray spectrum of an AGN was first observed in MR 2251--178 
\citet{b14}. The spectral complexity was due to absorption by 
partially ionized, optically thin, circumnuclear 
material. This material, termed the `warm absorber' was 
proposed as a common constituent of many AGN. ROSAT observations of 
nearby type~I AGN showed that warm absorption was not uncommon, 
see e.g. \citep{b28,b40}. Subsequent observations with 
\asca detected \oxyseven and \oxyeight absorption edges due to warm 
absorption in $\sim 50\%$ of type~I AGN (\citet{b35}, George 
\etal 1998). Observations in the UV band reveal warm absorption with 
multiple velocity components and ionization states, although the 
relationship between the UV absorbers and the X--ray absorbers 
is still not clear \citep{b8,b1}. The unprecedented spectral 
resolution of the gratings on board \chandra and \xmm has 
permitted the detection of discrete soft X--ray absorption features and 
emission lines for the first time. The resulting picture of the 
X--ray warm 
absorber in many type~I AGN is of an outflow exhibiting 
multiple narrow absorption lines corresponding to different 
ionization states e.g. \citep{b7,b21,b37,b15,b29,b42,b24}. The present 
generation of X-ray detectors also have the spectral resolution to detect 
unresolved transition arrays in moderately ionized Iron (Fe$^{0-15}$) in 
several AGN spectra \citep{b37, b2, b5, b33}.

Our picture of the X--ray warm absorber is now more complicated, and many
fundamental questions remain. Does the warm absorber `know' about the SBH 
mass, e.g. \citet{b26}, the accretion rate, or 
the AGN luminosity? 
Does radiation pressure dominate the warm absorber outflow? Is there 
a link between the warm absorber and the Fe~K band emission, e.g. Matt \etal
 (1994)? Only via 
the analysis of the soft X--ray spectra of a sample of 
type~I AGN can we answer such questions. Here we present the results of a 
uniform analysis of the soft X--ray data from a sample of fifteen 
type~I AGN observed with the high energy transmission grating spectrometer 
(\hetg) on board \chandra \citep{b22}. Our aim is to 
study the soft X--ray spectra of these AGN so that we may begin 
to answer some of the outstanding questions about warm absorption in AGN. Of 
course our study is limited by both our choice of sample and X-ray instrument.
 The \chandra \hetg bandpass is less sensitive to low ionization state 
absorbers than the \chandra \letg bandpass for example. 
Both the \letg and the RGS (aboard \xmmp) have a higher effective
area than the MEG at low energies but the \hetg spectral resolution
is superior to both the \letg and RGS. Utilizing the best spectral
resolution currently available is the prinicipal driving factor
for using the \hetg in the present study. There are many more
\hetg observations than \letg observations and the RGS bandpass
does not extend to the Fe K region of AGN spectra.
We note that in a uniform 
analysis, individual source pecularities (e.g. in the continuum modelling) may
 be missed. An additional complication is that the absorbers in some AGN are 
known to vary, so drawing general conclusions based on a snapshot of a 
variable absorber may not be warranted in some cases. Nevertheless, a uniform 
analysis is useful since variations between analysis software and/or 
methodologies can account for considerable differences in interpretation, 
often over the same data.

\section{The Sample and Data Analysis}
\label{sec:obs}

Our study is based on the sample of fifteen type~I AGN selected by 
\citet{b43}. The  AGN are listed in Table~\ref{tab:sample} were 
originally assembled for a study of the Fe~K band emission and had $z<0.05$, 
a total, first--order, high energy grating count--rate of 
$>0.05$ ct/s and the observations were in the \chandra public data archive 
\footnote{http://cda.harvard.edu/chaser/mainEntry.do} as 
of July 1, 2003. This constitutes a rather heterogeneous sample
that is not based on a scientifically motivated selection criterion.
Therefore, certain results and conclusions pertaining to {\it sample}
properties (such as the fraction of AGN exhibiting
signatures of photoionized outflows)
must be interpreted with the appropriate caution. 
However, members of the sample do satisfy one very important 
criterion relevant for this study, namely that these sources
are some of the brightest members of their class and therefore
lend themselves to performing
detailed X-ray spectroscopy with the \chandra HETGS. This is
not a coincidence because generally speaking, the highest
signal-to-noise members of a class tend to get accepted first by
selection panels for observations in the early years of a new
mission. 
 
We note that several of the AGN listed in 
Table~\ref{tab:sample} have been 
observed with \chandra again (for which that data
became public after July 1, 2003). 
Furthermore, an
additional thirteen AGN have been observed with the \chandra gratings 
that would fulfill the selection criteria of \citet{b43}. In the 
future we intend to extend our sample study to include these AGN \& more 
recent 
observations of the AGN in our sample. Also listed in Table~\ref{tab:sample} 
are the AGN redshifts 
(from NED\footnote{http://nedwww.ipac.caltech.edu} using 
21cm H{\sc~i} radiation measurements where possible), the RA and 
DEC (also from NED), the Galactic column density \citep{b10, b27}
 and the total exposure times of the spectra.

The \chandra data were reprocessed using {\tt ciao 2.1.3} and {\tt
CALDB} version 2.7, according to recipes described in {\tt ciao 2.1.3}
threads\footnote{http://cxc.harvard.edu/ciao/threads}. The
 instrument in the focal plane of \chandra during the observations was
         the \emph{HETGS}, which consists of two grating assemblies, a
   high--energy grating (HEG) and a medium--energy grating (MEG). Only
      the summed, negative and positive, first--order \chandra grating
spectra were used in our analysis.  The HEG bandpass is $\sim$0.8--10 keV and
         the MEG bandpass is $\sim$0.5--10 keV but the effective area of both
instruments falls off rapidly at either end of the bandpass. Since the
 MEG soft X-ray response is much better than the HEG we used the
  MEG as the primary instrument. 

We made effective area files (ARFs or
   \emph{ancillary response files}), photon spectra and counts spectra
     following the method of \citet{b42}.  We did not subtract
 detector or X-ray background since it is such a small fraction of the
  observed counts. For spectra with zero or few counts per bin
anywhere in the bandpass, attempting to subtract background whilst retaining
the best spectral resolution possible can result in worse systematic
errors compared to the case when no background subtraction is attempted. 
This is especially true when the background itself is weak, having
zero counts for most spectral bins.
However, background could be a source of contamination 
at the lowest energies of the MEG spectra
(where the effective area is smallest), 
for weak and/or heavily absorbed sources. We examined spectra taken from two 
strips, either side of the on-source data, to check the level of background 
for each data set. We found that contamination could be a problem in 
NGC~3516, NGC~3227 and NGC~7314, where the background level becomes 
comparable to the source intensity below $\sim 0.6$ keV. We will bear this 
in mind when interpreting the data. We note that five of the fifteen AGN 
in our sample were observed with the \chandra low energy transmission
 grating spectrometer 
(\letg). The high resolution camera (HRC) was used in four of these 
five \letg observations (as opposed to ACIS). 
However, order separation is not possible with 
the HRC and the properties of the \letg are considerably different to those 
of the \hetg (e.g. \letg has a lower spectral
resolution of only 0.05\AA\ FWHM ) and 
so inclusion of the \letg data would have complicated efforts to perform a 
uniform analysis and comparison amongst the sources. Note that the 
observations of three of the AGN (NGC~3516, NGC~3783 and MCG~$-$6-30-15) were
not made in a single observing period. NGC~3516 was observed in three parts, 
as detailed in Table~\ref{tab:sample}. However, two of the observations found NGC~3516 in a 
low flux state. So, in our analysis of the warm absorber in this AGN, we 
summed two of the observations into a single `low state' spectrum and 
compared this with the `high state' spectrum in the third observation. 
NGC~3783 was observed in five snapshots over a period of 
$\sim 124$ days. However, we analyze only the summed data in the present 
paper, since the spectral variability, which has been studied in detail by 
\citet{b29} is not great. The observation of MCG~$-$6-30-15 has a single 
sequence number and observation ID, but was made in three parts, spanning 
a period of $\sim 138$ days. The S/N in the individual parts was not 
sufficient to warrant analyzing the three parts separately so here we analyze
 only the summed data.

For comparison with some of the results from the \chandra observations, we 
studied \xmm spectra from some of the AGN in our sample. We used observations 
made with the \xmm EPIC PN instrument. The data used were obtained from the 
reduced data products from the {\it Xassist} database 
\footnote{{\tt http://xassist.pha.jhu.edu}}. AGN in our \hetg sample with 
corresponding \xmm observations are listed in Table~\ref{tab:sample}.

We treated the statistical
     errors on both the photon and counts spectra with particular care
       since the lowest and highest energies of interest can be in the
    Poisson regime, with spectral bins often containing a few, or even
   zero counts. When plotting data, we assign statistical upper and 
lower errors of $1.0 + \sqrt{(N+0.75)}$ and 
$N(1.0-[1.0-(1/(9N))-(1/(3\sqrt{N})]^3)$ respectively \citep{b12} 
on the number of photons, $N$, in a given spectral bin. When fitting 
the \chandra data, we used the $C$--statistic \citep{b6} for finding 
the best-fitting model parameters, and quote 90$\%$ confidence, 
one-parameter statistical errors unless otherwise stated. 
The $C$--statistic minimization algorithm is inherently Poissonian and 
so makes no use of the errors on the counts in the spectral bins 
described above. All model parameters will be referred to the source frame, 
unless otherwise noted. Note that since all models were fitted by first 
folding through the instrument response before comparing to the data, the 
derived model parameters \emph{do not} need to be corrected for 
instrumental response.

  We used XSPEC v.11.3.1 for spectral fitting to the \hetg spectra. All
spectral fitting was done in the 0.5--5~keV energy band, excluding the
 2.0--2.5 keV region, which suffers from systematics as large as $\sim
  20\%$ in the effective area due to limitations in the calibration of
                                                   the X-ray telescope
\footnote{http://cxc.harvard.edu/cal/cal\_present\_status.html}. 
We performed spectral fits using data binned at $\sim 0.02$\AA (which 
is close to the MEG FWHM spectral resolution  of $0.023$\AA), unless 
otherwise stated. There has been a continuous degradation of the quantum 
efficiency 
of \chandra ACIS with time, due to molecular contamination
\footnote{http://cxc.harvard.edu/cal/Acis/Cal\_prods/qeDeg/index.html}. 
In analyses of individual sources 
\citep{b42,b24,b25} we found that a pure ACIS correction
 (the worst--case effect) affects only the inferred intrinsic continuum 
(at less than $\sim 0.7$ keV) and 
does not affect the important physical parameters of 
warm absorber models. 
A detailed comparison of spectral results with and without
the ACIS degradation correction was given in \citep{b24} for the
case of NGC~4593. 
We also note that the data in our sample are from 
early in the \chandra mission when the ACIS degradation was not so severe
and the effects of the degradation are mitigated by the limited
signal-to-noise ratio of the data at the lowest energies. 

\section{The overall soft X--ray spectra}
\label{sec:soft}

The
 soft X-ray continuua of Type~I AGN can be quite complex. An excess 
over a simple power--law continuum model, as well
as possible absorption features, are common.
In our sample, nine of the fifteen AGN in 
Table~\ref{tab:sample} 
show features characteristic of a soft excess and/or absorption edges. 
NGC~3227, NGC~3516, IC 4329A and NGC~7314 are heavily absorbed in the soft 
X--ray band. F9, 3C~120 and Mkn~279 have relatively simple continuua,
 since their spectra show little or no evidence for a soft excess. Since the 
soft excess typically appears only in the 0.5--0.7 keV band of our data, we do
 not have enough information to
  constrain its origin and so sophisticated modelling of the continuum
     (such as with a power-law plus a blackbody model component) is not
warranted. We found that a broken power-law (although likely to be 
non-physical) is adequate to describe the intrinsic continuum
including any soft excess in the \hetg energy band. When 
partially covering cold absorption is added to the model a useful
empirical description of the overall spectra is obtained, leaving
residuals mostly due to complex, ionized absorption, and in some
cases some emission lines. Table~2 shows the best-fitting parameters
obtained with this model for each source, as well as observed
0.5--2~keV fluxes.

\section{Photoionization modeling of the AGN warm absorbers}
\label{sec:modeling}
We used the publicly available photoionization code XSTAR 2.1.kn3 
\footnote{http:heasarc.gsfc.nasa.gov/docs/software/xstar/xstar.html} to 
generate several grids of models of emission and absorption from 
photoionized gas in order to directly compare with the data. Version 2.1.kn3
 includes unresolved transition arrays (UTAs) of inner-shell transitions of 
Fe for the first time in XSTAR. It is important to include UTAs in our photoionized 
model grids since they provide the key to studying low ionization 
state absorbers in AGN and UTAs of moderately ionized Fe$^{0-15+}$ have 
been observed in the spectra of several type~I AGN (see e.g. 
\citep{b37, b33, b2, b5}).We used the default solar abundances in XSTAR 
(e.g. see Table~2 in \citet{b42}). 
Note that the XSTAR line database has 
been constructed from lines with published wavelengths, such as the
 Chianti line compilation 
\footnote{http://wwwsolar.nrl.navy.mil/chianti.html}.  This almost 
completely excludes the lines listed in \citet{b3} for 
example, which are based on HULLAC calculations and which are only 
reliable at about the 1$\%$ level.  These lines, due to L-shell transitions in
 Ne, Mg, Al, Si, S, Ar, Ca and Fe, are present in 
many Chandra grating spectra (see e.g. \citet{b29}), but so far 
they have mostly been omitted from the XSTAR database rather than 
compromise the accuracy of the database \citet{b16}.  

There may be additional absorption in most of these AGN, 
beyond the simple photoionized warm absorber models that we consider 
in this uniform analysis. Candidates for the extra absorption include 
neutral dust (e.g. possibly in the form of FeO$_{2}$ in MCG--6--30--15, 
see \citet{b21}) or deeper \oxyseven or \oxyeight edges due to Oxygen 
overabundance. Alternatively, a more complex model of 
the continuum may be more appropriate, incorporating relativistically 
broadened emission lines in the soft X-ray band (see e.g. 
\citep{b4,b23,b38}). However, investigation of these 
issues is beyond the scope of the present uniform analysis and we shall 
return to these important points in future work. 

The best spectral resolution of the MEG cannot {\it directly} constrain
turbulent velocities with a ``b-value'' of less than $170$ km $\rm{s}^{-1}$.
Furthermore, the XSTAR model spectra are calculated by XSTAR
on a grid that does not always preserve
accuracy in the line widths and line equivalent widths (Kallman,
private communication). Therefore, direct fits to the data using the
XSTAR spectra are approximate in the regions containing absorption
lines. However, the ionic column densities output by the XSTAR
are more accurate than the XSTAR spectra and can be used
for more detailed modeling of individual absorption lines. 
This level of detail is beyond the scope of the present
uniform analysis.
We used a velocity turbulence ($b$-value) of $170$ km $\rm{s}^{-1}$,
which corresponds approximately to the limiting MEG spectral resolution 
($\sim 300$ km $\rm{s}^{-1}$ FWHM at 0.5 keV), given that the data  
cannot directly constrain smaller line widths.
It is possible to constrain $b$ using a curve-of-growth analysis
if one has several absorption line measurements, and this has
been done for some of the sources in our sample in more
detailed studies. For example, in Mkn~509 $b$ is consistent
with $\sim 100$ km $\rm{s}^{-1}$ (\citet{b42}), with similar
results obtained for other sources. Since $b$ is generally
comparable to or less than the spectral resolution, our 
adopted value of $b$ is justified for the purpose of spectral
fitting.

\begin{figure}
\begin{center}
\includegraphics[height=3.35in,width=3.35in]{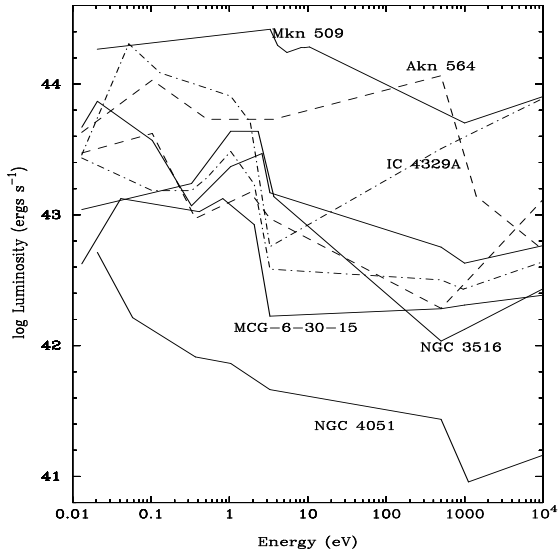}
\caption{Spectral energy distributions (SEDs) used for photoionization 
modelling to the \chandra data of ten AGN in our sample (see 
\S\ref{sec:sed}). In order of decreasing luminosity at 10keV 
(right-hand-side of graph), the SEDs correspond to
Mkn 509, IC 4329A, NGC 5548, NGC 3783, Mkn 766, Akn 564, NGC 4593, 
NGC 3516, MCG-6-30-15 and NGC 4051 respectively.}
\label{fig:sed}
\end{center}
\end{figure}

\subsection{XSTAR modelling procedure}
\label{sec:sed}

 First, we constructed a spectral energy distribution (SED) for each AGN
       in our sample, according to the method described in \citet{b42}. 
In principle, derivation of the SED should be an iterative
process since the intrinsic X-ray spectrum is obtained from
fitting the data using an SED that must already contain information
about the X-ray spectrum. In practice, the method employed
(described in \cite{b42}) is a two-step process in which
the first step is an
initial estimate of the intrinsic X-ray spectrum is obtained
by fitting the X-ray data with a power-law or broken power-law
only and no photoionized absorber. The resulting estimated SED
is used to generate grids of photoionization models that
are used to obtain better fits to the data and to
derive a new intrinsic X-ray spectrum. This is then used to generate
another SED which is in turn used to generate a new set of photoionzation
model grids and the latter are the ones used for the final model fits.
Average radio, infrared (IR), optical and ultraviolet (UV)
  fluxes  were obtained from \citet{b41} and NED. Note that flux
measurements, at different wavelengths, obtained from historical data  
(\citet{b41}, NED) are
    not in general contemporaneous. For each source,
the 0.5 keV intrinsic model flux from the intrinsic X-ray
continuum model was then simply joined onto the last point
of the UV part of the SED by a straight line in  log-log
 space. The hard X-ray power law was extended out to 500 keV. 
Generally, the high-energy
cut-off in the observed X-ray spectra of Seyfert galaxies lies in
the range ~100-500~keV; only blazars and BL Lacs are observed
to have significant flux beyond 500 keV. The ionization balance is
is not sensitive to the exact position of the cut-off between
100-500 keV.
At the low-energy end, it has been shown (for example, by
\cite{b90}) that if the observed IR continuum
carries a significant thermal component from reprocessing 
of the instrinsic continuum, using the observed SED as an
input to photoionization models can potentially affect the
resulting ionization balance of the plasma. However, our
model fits are driven by X-ray features and we showed in a
detailed study of NGC~4593 (\cite{b24}) that removing the
prominent IR-optical
continuum bump from the SED  yielded warm absorber parameters
that were within the 90\% confidence intervals obtained when
the bump was not removed.
Figure~\ref{fig:sed} shows the 
SEDs that we used for the ten AGN in our sample that exhibited warm 
absorption (see \S\ref{sec:ten} below).

The photoionization model grids used here are two-dimensional,
  corresponding to a range in values of total neutral Hydrogen column density,
  $N_{\rm H}$, and the ionization parameter, $\xi = L_{\rm ion}/(n_{e}
R^{2})$.   Here  $L_{\rm ion}$ is the ionizing luminosity in the range
    1--1000 Rydbergs, $n_{e}$ is the electron density and $R$ is the
 distance of the illuminated gas from the ionizing source.  
The ionizing luminosities were calculated from the SEDs by
normalizing the absorption-corrected 0.5--2~keV fluxes.  The grids
  were computed for equi-spaced intervals in the logarithms of $N_{\rm
        H}$ and  $\log{\xi}$, in the ranges $10^{19}$ to $5 \times 10^{23} 
{\rm cm^{-2}}$ and $-1.0$ to $+4.0$ erg  cm ${\rm s^{-1}}$ respectively. 
We computed grids with $n_{e}$ in the range $10^{2-11} \ {\rm cm^{-3}}$. 
For XSTAR models of the \emph{absorber}, we confirmed that 
results from fitting the photoionization models to the X-ray data were
     indistinguishable for densities in the range $n_{e}=10^{2} \ {\rm
cm^{-3}}$ to $10^{11} \ {\rm cm^{-3}}$ for all of the AGN. Hereafter 
we will use $n_{e}=10^{8} \ {\rm{cm}^{-3}}$  unless otherwise stated. 

Our aim was to fit the MEG spectra with a simple power-law or a 
broken power law continuum (whichever was the better fit), 
modified by absorption from photoionized gas (derived from the AGN SED), 
neutral gas intrinsic to the AGN and Galactic absorption. All model 
fitting to the data was carried out using XSPEC v11.3.1. In order to fit 
XSTAR photoionization 
model grids to the data, we required an offset velocity for the 
warm absorber for each of the AGN. Therefore, first we tested 
for the following He-like and H-like absorption lines in each spectrum within 
$\sim \pm 5000$ km $\rm{s}^{-1}$ of the line rest--energy in the AGN 
frame: \nlya ($\lambda 24.781$ \AA), \oxysevenr($\lambda 21.602$ \AA), 
\oxla($\lambda 18.969$ \AA), \neniner($\lambda 13.447$ \AA), 
\nelya($\lambda 12.134$ \AA), \mgelevenr($\lambda 9.169$ \AA), 
\mglya($\lambda 8.421$ \AA), \sithirteenr($\lambda 6.648$ \AA), 
\silya($\lambda 6.182$ \AA). Note that for individual 
absorption features, the statistical errors on the line energies are 
typically very small and so the systematic uncertainty likely 
dominates these errors. The relative wavelength accuracy of the MEG is 
$0.0055$ \AA\ and the absolute wavelength accuracy of the MEG is $0.011$ 
\AA \footnote{http://asc.harvard.edu/proposer/POG/html/HETG.html}.

Next, we established an offset velocity for a particular AGN by finding 
the weighted--mean offset velocity (and the 90$\%$ confidence limits) 
of the centroids of all the absorption 
lines detected at $>90\%$ significance in the spectrum. Some AGN exhibited 
groups of two or more absorption features separated from another such 
group by more than the 90$\%$ errors on their respective centroid 
velocities. In 
these cases, we established more than one weighted--mean offset velocity. 
Finally, we fitted XSTAR models only to those AGN where two or more 
prominent absorption signatures had similar offset velocities (within 90$\%$ 
confidence limits). This approach has the merit of being uniform, but it 
runs the risk of being too conservative and we may be biasing our study in 
favour of more prominent warm absorption signatures. On the other hand, our 
method \emph{does} have the merit of ignoring statistically spurious features.
 We tested for the presence of multiple ionization components in the absorbers 
by investigating the statistical significance of additional model components 
within the 90$\%$ confidence limits of the weighted-mean offset velocity.

\begin{table*}
 \begin{minipage}{145mm}
   \caption{Best-fitting simple broken power-law continuum parameters 
(including neutral absorption with column density $N_{H}$, covering
a fraction, CF, of the source), and observed 0.5--2~keV fluxes. 
The broken power-law has photon indices $\Gamma_{1}$ and $\Gamma_{2}$,
and a break energy $E_{B}$.
}
   \begin{tabular}{@{}lrrrrrr@{}} 
   \hline
Source & cold $N_{H}$ & CF & $\Gamma_{1}$ & $E_{B}$ & $\Gamma_{2}$ & Flux (0.5-2.0keV)\\
 &($10^{20} \rm{cm}^{-2}$) & & & (keV) & & ($10^{-12}$ erg $\rm{cm}^{-2} \rm{s}^{-1}$)\\
\hline
F9 & $<1$ & &$2.10^{+0.09}_{-0.07}$  &$1.06^{+0.08}_{-0.04}$ & $1.78^{+0.02}_{-0.03}$ & 11.9\\
3C~120 & $<1$& & $1.79^{+0.11}_{-0.10}$ & $0.99^{+0.29}_{-0.16}$& $1.68\pm 
0.03$ & 16.8\\
NGC~3227 & $188\pm24$ & $0.84^{+0.04}_{-0.05}$& & & $1.32^{+0.19}_{-0.18}$& 1.6\\
NGC~3516 (low)& $247^{+15}_{-17}$ &$0.82^{+0.04}_{-0.03}$& & & 
$1.65^{+0.10}_{-0.12}$ & 2.0\\
NGC~3516 (high) & $213^{+21}_{-19}$&$0.79^{+0.02}_{-0.03}$ & & &
$1.92^{+0.11}_{-0.13}$ & 3.7\\
NGC 3783 &$102^{+1}_{-2}$&$0.75\pm 0.01$& $3.07^{+0.05}_{-0.06}$& $1.23\pm 0.01$ & $1.92^{+0.02}_{-0.03}$ & 11.8\\ 
NGC 4051 & $<1$ & & $3.08^{+0.14}_{-0.07}$&$1.23^{+0.03}_{-0.09}$ &$1.77\pm 0.05$ & 12.6\\ 
Mkn 766 & $2\pm 1$& & $3.10^{+0.16}_{-0.28}$& $0.91^{+0.08}_{-0.04}$&$1.85^{+0.03}_{-0.04}$& 15.0\\ 
NGC 4593 & $3^{+2}_{-1}$& & $2.72^{+0.19}_{-0.16}$&$0.96^{+0.05}_{-0.02}$&$1.80\pm0.04$& 20.9\\ 
MCG-6-30-15 & $6\pm2$& & $3.82^{+0.24}_{-0.21}$&$1.02^{+0.04}_{-0.02}$&$1.76^{+0.04}_{-0.05}$& 9.7\\ 
IC 4329A & $32^{+4}_{-3}$& & $3.66^{+0.28}_{-0.22}$&$0.98\pm 0.03$ & $1.83^{+0.02}_{-0.03}$& 77.9\\ 
Mkn~279 & $<1$ & & $1.96^{+0.17}_{-0.16}$& $0.89^{+0.14}_{-0.09}$& 
$1.56\pm0.03$ & 5.6\\
NGC 5548 & $<1$ & & $2.80^{+0.16}_{-0.13}$ & $0.95\pm 0.03$ & $1.46^{+0.04}_{-0.03}$ & 10.0\\ 
Mkn 509 & $<1$ & & $2.50^{+0.08}_{-0.09}$ & $0.96^{+0.09}_{-0.06}$& $1.63\pm 0.03$ & 23.1\\ 
NGC~7314 & $107\pm5$ & $0.97\pm 0.01$& & & $2.00^{+0.06}_{-0.05}$ & 4.9\\
Akn 564 & $4^{+2}_{-1}$& & $3.16^{+0.11}_{-0.15}$& $1.43^{+0.06}_{-0.05}$& $2.48\pm 0.05$ & 44.9\\
\hline
\end{tabular}
\end{minipage}
\end{table*}

\subsection{XSTAR model results}
\label{sec:ten}

\begin{figure*}
\includegraphics[height=8in,width=7.0in]{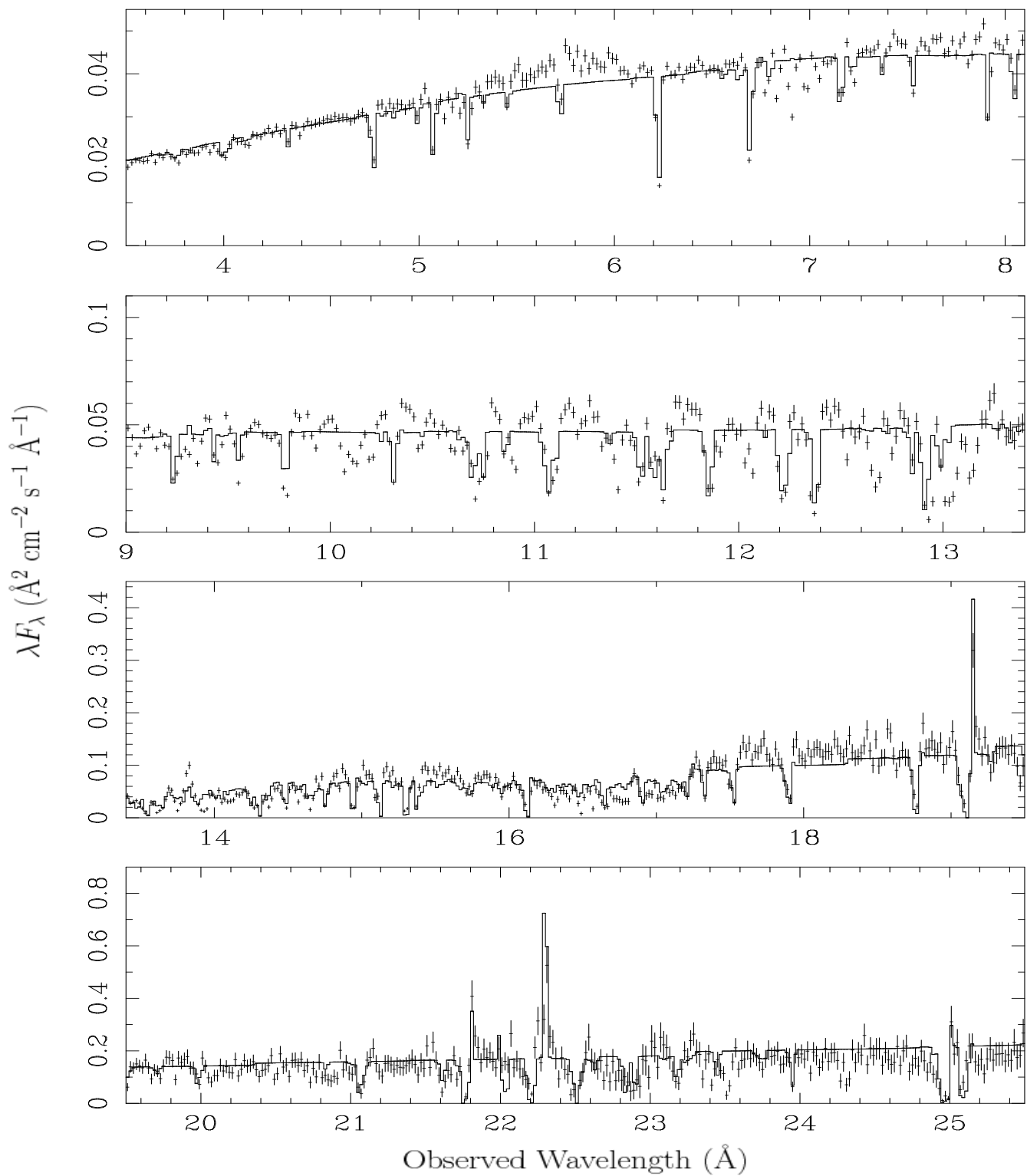}
\caption{MEG observed photon spectrum compared to the 
best--fitting simple photoionized absorber plus emitter model (solid line) in 
the energy range 0.5-3.0 keV for NGC 3783 (see Tables~\ref{tab:1wabs} and 
\ref{tab:em}). The data were binned at 0.02\AA\ for 
clarity. There are 
considerable gaps in the XSTAR database of Fe transitions 
so multiple Fe absorption features are not accounted for by the 
model. As can be seen, the absence of these lines from the model fit serves to 
alter the continuum fit in several places. There are also known
problems with the fit. The continuum fit is poor around 2.0-2.5 keV ($\sim$ 5-6 
$\AA$) in general due to calibration uncertainties. The continuum 
excess relative to the model around $\sim 0.8$ keV ($\sim$ 13-15 $\AA$) was 
also found by N03, using a different photoionization code (ION) to model the 
data. }
\label{fig:3783}
\end{figure*}

\begin{figure*}
\includegraphics[height=8in,width=7in]{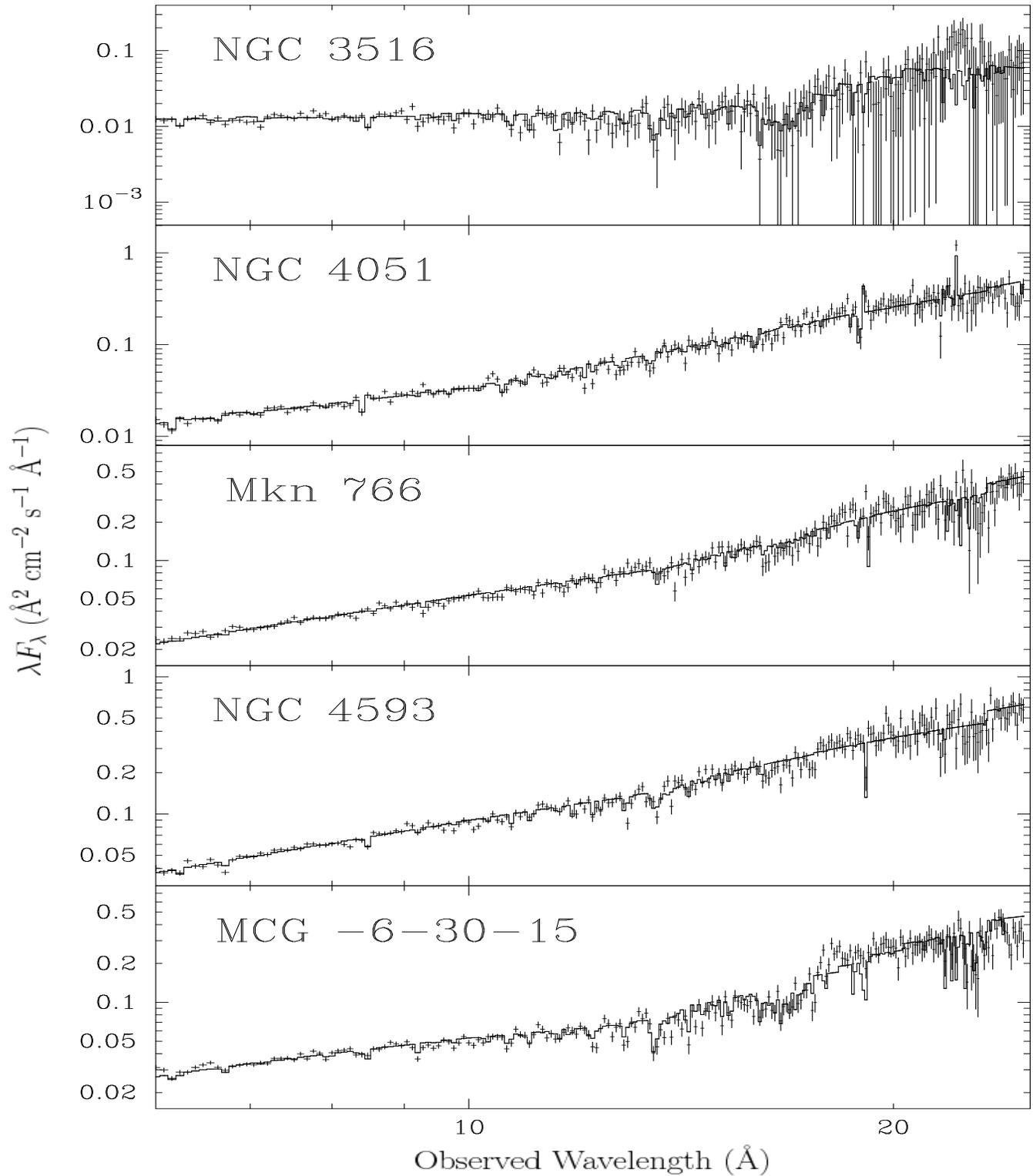}
\caption{MEG observed photon spectra compared to the 
best--fitting simple photoionized absorber plus emitter models (solid line) 
for five of the AGN in our sample (see
Tables~\ref{tab:1wabs} and \ref{tab:em}).
The data were 
binned at 0.08\AA\ for clarity in the plots
(but  0.02\AA\ for fitting).
The plot for NGC~3516 is that for the higher flux state,
which is the one that
showed evidence for a Fe UTA. Note that the emission feature at 
$\sim 0.55$ keV (observed frame) in NGC~3516 is due to the background, not 
the source (see text).}
\label{fig:ufspec1}
\end{figure*}

\begin{figure*}
\includegraphics[height=8in,width=7in]{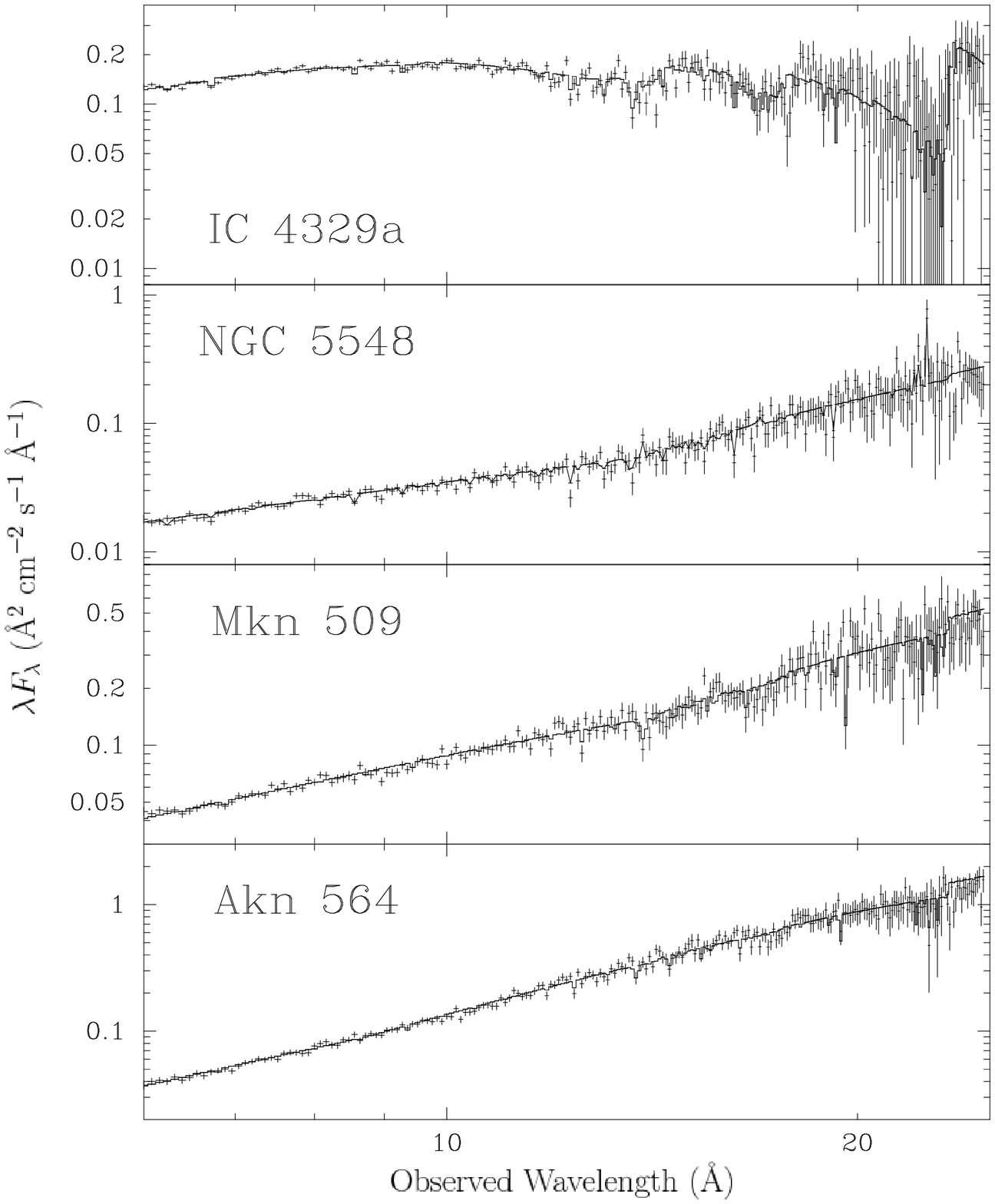}
\caption{As for Fig.~3 for four more of the AGN in our sample.}
\label{fig:ufspec2}
\end{figure*}

Table~\ref{tab:1wabs} lists the best--fit warm absorber models that we 
established for our AGN spectra. Figures~\ref{fig:3783},\ref{fig:ufspec1} 
and \ref{fig:ufspec2} show the best--fit models overlaid on the \hetg spectra.
We note that in Table~\ref{tab:1wabs}
the derived column densities of each warm
absorber component do not depend on the ``b-value'' assumed in the
XSTAR models because the absorption lines are unresolved. Also, these column
densities are not soley driven by the absorption lines.
In particular, in the case of NGC~3783, altough our simplistic model
gives a poor fit in some regions of the spectrum, missing some 
absorption lines altogether, the column densities of the four
warm absorber components are not sensitive to this since
the four numbers are determined by a fit to tens of absorption lines
as well as to the overall spectral shape.

Only ten of the fifteen AGN in our sample satisfied our criteria for
 selection and spectral fitting with XSTAR grids. The AGN F9, 3C 120, Mkn~279,
 NGC~3227 and NGC~7314 are not included in Table~\ref{tab:1wabs}. In the case 
of F9, we found weak \oxla and \oxysevenr absorption signatures 
($\leq 90\%$ confidence). 3C~120 exhibited 
strong \oxla absorption but no other absorption signatures at $> 90 \%$ 
confidence. In Mkn~279, \nelya was the only absorption feature present at 
$>90\%$ confidence. In 
NGC~3227 and NGC~7314, the soft X--ray band was very strongly absorbed, so 
we could not distinguish between a cold absorber and a mildly warm absorber.
In NGC~3227 we only detected a single statistically significant absorption 
feature (\silya at $\sim -1700$ km $\rm{s}^{-1}$
relative to systemtic velocity). Note that previous, 
higher S/N data of the  heavily absorbed AGN in our sample (NGC~3227 \& 
NGC~7314) have in some cases detected evidence for photoionized absorption 
(see e.g.\citet{b35,b93,b13}). Spectral plots of the data for the
five AGN that are not shown in Figure~\ref{fig:ufspec1} can be found
in the HETG public database, {\it HotGAS}\footnote{{\tt http://hotgas.pha.jhu.edu}}.

\begin{table*}
 \begin{minipage}{135mm}
  \caption{Detailed spectral fitting results for Photoionization models 
\label{tab:1wabs}. Column 2 shows the best--fit cold absorption. Column 5 
shows the 
mean offset velocity (negative=blueshift w.r.t. systemic) for the warm 
absorber components. Column 6 shows rate of mass 
outflow in terms of $C$, the product of the 
filling factor \& the covering factor. Column 7 shows the fit--statistic 
(which includes emission model components from Table~\ref{tab:em}). 
Column 8 shows the increase in
 fit-statistic when the absorber component (3 dof) is removed from the model 
fit and the model is re-fitted . 
$^{c}$ The true 90$\%$ confidence upper or
 lower limit could not be determined in these cases. The effective range of 
the offset velocity should be taken to lie between the best-fit value 
and the given 
upper or lower limit.}
\begin{tabular}{@{}lrrrrrrr@{}}
\hline
Source & $N_{H}$ &$N_{wabs}$ 
&$\log \xi$ &$v_{wabs}$ & $\dot{M}_{\rm out}/C$ &C-stat 
& $\Delta$ C\\
               &($10^{20} \rm{cm}^{-2}$) &($10^{21}
         \rm{cm}^{-2}$) &ergs cm $\rm{s}^{-1}$ &km
$\rm{s}^{-1}$ & $M_{\odot}$/yr & & \\
\hline
NGC 3516 (low)&$250\pm 20$  
      &$3\pm1$ &$2.4^{+0.2}_{-0.1}$ &$-910^{+140}_{-155}$ &
                1.1  & 1363& 60\\ 
high flux & $210^{+20}_{-30}$
      &$16^{+8}_{-4}$ &$2.5\pm 0.1$ &$-585^{+75}_{-65}$ &
                 2.2 &1414& 131\\ 
 &
      &$2\pm1$ &$1.0^{+0.2}_{-0.1}$ &$-1685^{+135}_{-170}$ &
                   210 & & 18\\ 
\hline
NGC 3783 &$102^{+1}_{-2}$  
  &$20^{+7}_{-3}$ &$2.9\pm 0.1$ &$-505 \pm 15$ & 2.3& 9091 
& 689\\ 
   &&$6\pm 1$ &$2.1\pm 0.1$ & $-515 \pm 15$
& 14& & 5070\\ 
   & &$2^{+0.2}_{-0.1}$ &$0.4\pm 0.1$ &$-545^{+30}_{-20}$ 
& 720& & 1039\\ 
   &&$5^{+0.4}_{-1}$ &$3.0 \pm 0.1$ &$-1145^{+55}_{-30}$ 
&3.5 & & 696\\ 
\hline
NGC 4051
   &$<1$ &$0.1^{+0.1}_{-0.1}$ &$1.0 \pm 0.3$ &$-520^{+c}_{-90}$ 
& 2.5&1296& 29\\ 
   & &$0.9^{+0.5}_{-0.3}$ &$2.6^{+0.4}_{-0.1}$ &$-600^{+90}_{-65}$ 
& 0.1 &    &  76\\ 
   & &$90^{+40}_{-50}$ &$3.8\pm 0.1$ &$-2230^{+50}_{-60}$
& $<0.1$ &    &  41\\ 
\hline
Mkn 766 &$2\pm1$
      &$0.2^{+0.3}_{-0.1}$ &$<0.6$ &$-75^{+25}_{-70}$ &
                  310 &1289& 8\\ 
  &  &$0.6^{+1.6}_{-0.5}$
                          &$3.1^{+0.3}_{-0.1}$ & $-25^{+c}_{-145}$ & $<0.1$
& & 6\\ 
  &  &$0.2^{+0.3}_{-0.1}$
                          &$2.0\pm0.1$ & $-25^{+c}_{-25}$ & 0.4& 
& 41\\ 
\hline
NGC 4593 &$3^{+2}_{-1}$
       &$2\pm1$ &$2.4^{+0.1}_{-0.2}$ &$-95^{+c}_{-10}$&
0.3 & 1393& 25\\ 
   &  &$4^{+8}_{-2}$ &$3.3^{+0.2}_{-0.4}$ & $-95^{+c}_{-30}$
 & $<0.1$&  &43\\ 
\hline
MCG-6-30-15 &$6\pm2$ &$4 \pm 1$
                          &$0.2 \pm 0.1$ &$+30^{+c}_{-60}$& 9.1 &1673 & 
226 \\ 
  &  &$3\pm 1$
                          &$2.1\pm 0.1$ & $+15^{+c}_{-55}$& $<0.1$
& & 240 \\ 
  &  &$30^{+60}_{-20}$
                          &$3.7^{+0.1}_{-0.3}$ & $-1555^{+80}_{-130}$ &0.2 
&& 62\\ 
\hline
IC 4329a &$32^{+4}_{-3}$ &$2.3^{+0.2}_{-0.3}$
        &$0.2 \pm 0.1$ &$-100^{+c}_{-55}$& 750 &1448 & 162\\ 
  &  &$1.4 \pm0.3$ & $2.2\pm 0.1$& $-100^{+c}_{-20}$
& 7.8& & 175\\ 
\hline
NGC 5548
    &$<1$ &$0.6 \pm 0.2$ &$2.2\pm 0.2$ &$-560^{+c}_{-80}$& 89
&1065& 126\\ 
  & &$50\pm40$ & $3.9^{+0.1}_{-0.2}$& $-830^{+270}_{-c}$ &3.3&&105\\ 
\hline
Mkn 509 &$<1$ &$0.5\pm 0.2$ &$2.3^{+0.2}_{-0.1}$
  &$-140^{+c}_{-40}$& 14& 1137 & 36\\ 
   & &$0.1\pm 0.1$ &$0.6^{+0.8}_{-0.3}$ & $-140^{+c}_{-100}$
 & 1100& & 27\\ 
\hline
Akn 564 &$4^{+1}_{-2}$
             &$0.1\pm 0.1$ &$<0.4$ &$-140^{+c}_{-50}$&
                 930  &1354& 18\\
  &  &$0.2^{+0.3}_{-0.1}$
                          &$2.6\pm 0.2$ & $-140^{+c}_{-15}$ &5.4 && 
131\\ 
\hline
\end{tabular}
\end{minipage}
\end{table*}

Table~\ref{tab:compare} shows results from the literature for model fits to 
X-ray spectra from the AGN in our sample. In general, given the limitations of
 our methodology, the agreement is actually quite good. Our uniform analysis 
ignores individual continuum peculiarities in individual AGN and our 
selection criteria for absorbers are more conservative than many in the 
literature. In spite of this, our method seems to pick up the same significant 
absorber components in the literature. Perhaps the best example of this can be
 seen by comparing our results from NGC~3783 in Table~\ref{tab:1wabs} with 
those from a detailed analysis by \citet{b29} of the best S/N X-ray spectrum 
yet. Table~\ref{tab:compare} shows that \citet{b29} find two velocity 
components of the warm absorber with three ionization states. Our simple, 
uniform method picks out the same velocity components at $\sim \ -500$, 
$-1100 \rm \ km \ s^{-1}$
and reveals the three ionization states, although we do not establish 
high velocity components for the middle 
and low ionization states. The total column density of the absorbers agrees 
very well with the 
value of $\sim 3 \times 10^{22} \rm{cm}^{-2}$ found by \citet{b29}.
Our simple, uniform analysis means that our continuum fit is not as good as 
that found by \citet{b29}, but it is encouraging that we reproduce most of 
their model parameters, given that their analysis of this spectrum was far 
more detailed than ours.

Of course, there are AGN for which 
individual source peculiarities mean that our 
results agree less well with those in the literature. In the case of \mcg, 
\citet{b21} and \citet{b38} come to very different conclusions concerning 
both the origin of the continuum in this AGN \emph{and} the parameters 
associated with the warm absorber. 

\begin{table*}
 \begin{minipage}{160mm}
  \caption{Results from the literature for the spectra of the type~I AGN in 
our study. For brevity, we only
 include the larger studies that parameterized $N_{H},\log \xi$
 and/or the velocity of the warm absorber. We abbreviated X-ray instruments
 as: XR=\xmm RGS, XE=\xmm EPIC-pn, CL=\chandra \letg, CH=\chandra \hetg,
 BX=\bsax, RX=\rxte. Gaps indicate value of $\log \xi$ \emph{not} specified 
in the 
literature. In these cases, either no photoionization modelling was carried 
out (Mkn 766, NGC 4051, NGC 3516) or a different form of the ionization 
parameter was specified (U or $U_{ox}$), with insufficient information to 
translate to values of $\log \xi$. $^{a}$ Includes a model 
continuum component
 representing relativistically broadened \oxla or a radiative recombination
 continuum from material close to a Kerr BH. $^{b}$ Continuum model included
a column density of
 $3 \times 10^{17} \rm{cm}^{-2}$ of $\rm{FeO}_{2}$. $^{c}$ Continuum model
 included a blackbody with temperature of $kT=473$~eV and a neutral absorber 
with a column density of $\sim 3 \times 10^{21} \rm{cm}^{-2}$. $^{d}$ Used 
$z=0.01676$ from optical measurements rather than $z=0.01717$ from 21cm 
H~{\sc i} measurements. \label{tab:compare}.
}
\begin{tabular}{@{}lrrrrr@{}}
\hline
Source & Instrument & $N_{\rm wabs}$ & $\log{\xi}$ & velocity & Reference \\
 & & ($10^{21} \rm{cm}^{-2}$) & (erg cm $\rm{s}^{-1}$)& km $\rm{s}^{-1}$ & \\
\hline
NGC 3516 & XR & $\sim 8$ & & & \citet{b44} \\
	&CL & $\sim 8$ & & & \citet{b45} \\
	&XE,CL,BX & $16$ &  & $\sim -1100$ & \citet{b53} \\
	&	& $3-10$ &  & $\sim -1100$ & \\
\hline
NGC 3783 & CH&$20.0^{+11.6}_{-1.4}$ &  &[-400,-600],[-1000,-1300] & \citet{b29}\\
	&	&$10.0^{+4.1}_{-2.9}$&  &[-400,-600],[-1000,-1300] & \\
	&	&$7.9^{+2.1}_{-1.9}$&  &[-400,-600],[-1000,-1300] & \\
\hline
NGC 4051 & CH & $\sim 1$ &  & $-2340\pm 130$ & \citet{b7} \\
	 &	& $\sim 0.1$ &  & $-600 \pm 130$ & \\
	 & XE & $\sim 200$ & $\sim 3.8$ & $\sim -6500$ & \citet{b34}\\
	 &	& $\sim 6$ & $2.7 \pm 0.1$ & $0 \pm 200$ & \\
	 &	& $\sim 2$ & $1.4 \pm 0.1$ & $0 \pm 200$ & \\
	 & XE & [0.03,1] & [0,2.8] & $400\pm 100$ & \citet{b71}\\
\hline
Mkn 766 &CH & & &$0 \pm 160$ & \citet{b38}$^{a}$ \\ 
\hline
NGC 4593 & CH,RX& $5.4^{+1.5}_{-0.8}$ & $\sim 2.5$ & $-135\pm 40$ & \citet{b24}\\
	 & XR,CL & $1.6 \pm 0.4$ & $2.6 \pm 0.1$ & $-400\pm 120$ & \citet{b94}\\
	 &	& $0.1\pm 0.05$& $0.5 \pm 0.3$ & $-400\pm 120$ & \\
\hline
MCG-6-30-15 & CH & $\sim 32$ & $\sim 2.5$ & $\sim 0$ & \citet{b21}$^{b}$\\
	&	& $\sim 5$ & $\sim 0.7$ & $\sim 0$ & \\
	& XR & $\sim 2$ & $2-3$ & $-1900 \pm 140$ & \citet{b38}$^{a}$\\
	&    & $\sim 2$ & $0.5-2$ & $-150\pm 130$ & \\
\hline
IC 4329a & XR & $\sim 1.3$&$-1.4\pm 0.1$ &$\sim 0$ & \citet{b91}$^{c}$\\
	&     & $\sim 0.3$ & $0.6\pm 0.1$ &[-100,-300] & \\
	&     & $6.6\pm 0.4$ & $\sim 1.9$ &[-200,0] & \\
	&     & $2.0\pm 0.5$ & $2.7\pm 0.1$ & [-140,+180]& \\
\hline
NGC 5548 & CL,CH &$2.5\pm0.5$ & $2.3\pm 0.1$ & -530 & \citet{b39}$^{d}$ \\
	&     & $1.0 \pm 0.4$ & $1.9\pm 0.1$ & -530	& \\
	&     & $0.6^{+0.9}_{-0.4}$ & $-0.2\pm 0.2$ & -530	& \\
\hline
Mkn 509 & CH& $2.1^{+0.4}_{-0.5}$ & $1.8^{+0.1}_{-0.2}$ & $-200\pm 100$ & \citet{b42}\\
\hline
Akn 564 & CH  & $\sim 1$ & $\sim 2$ & $\sim -200$ & \citet{b48}\\
	&     & $\sim 1$ & $\sim 1$ & $\sim -200$ & \\
\hline
\end{tabular}
\end{minipage}
\end{table*}

Unresolved transition arrays (UTAs) of Fe are a key diagnostic of material 
in a low-ionization state, which could account for the bulk of the mass 
outflow from AGN (depending on the absorber 
density and geometry). The present generation of X-ray telescopes have the spectral resolution to 
detect blends of Fe inner-shell absorption transitions in the ionized 
outflows from AGN for the first time.  Unresolved arrays of inner--shell 
(2p-3d) transitions in moderately ionized Fe$^{0-15+}$ have been 
observed in the spectra of several type~I AGN (see e.g. \citet{b2} and 
references therein). 
\citet{b2} calculate Fe UTAs for an ionization parameter in the range 
$\log \xi \sim 0.1-1.1$. The UTA appears as a jagged, broad trough between 
$\sim 16-17$\AA . As $\xi$ increases, the centroid of the UTA trough shifts 
towards shorter wavelengths, the dominant contributions come 
from more highly ionized states of Fe and the Fe UTA broadens.

\begin{figure}
\includegraphics[height=4in,width=3in]{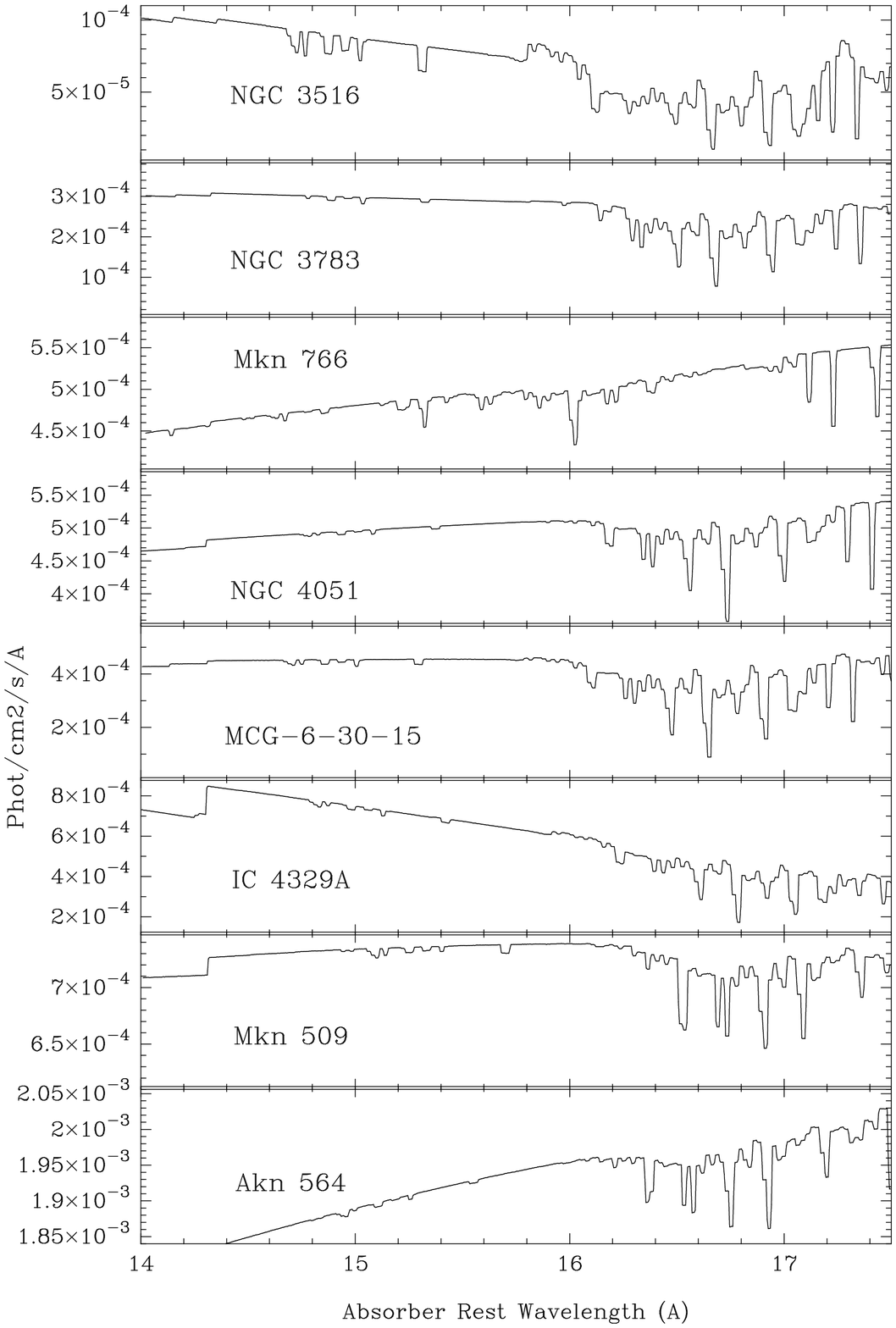}
\caption{Plots of the model Fe UTAs produced by warm absorber components from
 Table~\ref{tab:1wabs} with best-fitting ionization parameters in the range 
$\log \xi \sim 0.0-1.0$ (see text).}
\label{fig:uta}
\end{figure}

Of the ten AGN in our sample that exhibit signatures of absorption due to 
photoionized material, only two AGN (NGC~4593 and NGC~5548) 
did not exhibit statistically significant evidence for a low ionization 
(Fe UTA) absorber component. Note that in the case of NGC~3516, we only found 
a low ionization component at statistical significance in the high flux state
 of this AGN. Figure~\ref{fig:uta} shows the best-fit model 
of the UTA in each of the seven AGN with a warm absorber component where 
$\log \xi <1.1$ (see Table~\ref{tab:1wabs}). The wavelength scale 
in Figure~\ref{fig:uta} corresponds to the outflowing warm absorber 
rest-frame in each case.

\begin{table*}
 \begin{minipage}{80mm}
  \caption{Properties of Emitters \label{tab:em}. $^{a}$ We used 
$n_{e}= 10^{8} \rm{cm}^{-3}$ as a default electron density since the data 
could not discriminate between emission model grids in the range 
$10^{2-11}\rm{cm}^{-3}$.}
   \begin{tabular}{@{}lrrrr@{}}
   \hline
Source &$N_{em}$ &$\log \xi_{em}$ 
&$n_{e}^{a}$ &$v_{em}$\\
       &($10^{21} \rm{cm}^{-2}$) 
&(ergs cm/$\rm{s}$) 
&($\rm{cm}^{-3}$) 
&(km /$\rm{s}$)\\
\hline
NGC 3783 
  &$2.4^{+1.1}_{-1.3}$ &$1.61^{+0.15}_{-0.16}$ &$>280$ 
& $-5 \pm 45$ \\ 
NGC 4051 & $108^{+34}_{-88}$&$2.14^{+0.28}_{-0.40}$ & $>3.7 \times 10^{6}$
& $-160 \pm 75$ \\ 
NGC 5548 & $ 15.7^{+9.1}_{-13.6}$ &$1.40^{+0.15}_{-0.36}$ &$>230$&
$-115 \pm 125$\\ 
\hline
\end{tabular}
\end{minipage}
\end{table*}

\subsection{Soft X--ray emission in AGN}
\label{sec:emission}

As can be seen from Figs~\ref{fig:3783}, \ref{fig:ufspec1} and 
\ref{fig:ufspec2}, narrow 
emission lines were apparent in several of the AGN in our sample. The 
spectra of NGC~3783, NGC~4051 and NGC~5548 displayed the most prominent 
narrow lines. Narrow emission lines were less prominent in the 
spectra of Mkn~279, NGC~4593, Mkn~509 and MCG--6--30--15. The 
spectra of NGC~3227, NGC~3516 and NGC~7314 showed broad emission features 
around $\sim 0.55-0.6$ keV. However the broad 
emission features in the soft X--ray spectra of these three heavily absorbed 
AGN were present in the respective background spectra and are therefore 
not intrinsic to the AGN. We tested for \oxyseven and \nenine triplet 
line emission and 
\oxla and \nelya narrow lines in the spectra of the ten AGN listed above. 
We followed the fit procedure in \S\ref{sec:sed} except that
the Gaussian model component was not inverted. In a given AGN, we 
established the weighted--mean offset velocity of the centroids of emission 
lines detected at $>90\%$ significance in the spectrum. We then fitted 
XSTAR emission line model grids at the mean offset velocity to those AGN 
spectra displaying multiple ($>2$) emission signatures at $>99\%$ 
significance. 

From the definition of $\xi$, the emitting material lies at a distance 
$R \sim 3 \times 10^{-6} \sqrt{L_{\rm ion, 44}/(\xi_{100}n_{e})}$ pc from the radiation 
source, where $L_{\rm ion, 44}$ is the 1--1000 Ryd ionizing radiation from the 
AGN (in units of $10^{44}$ erg $\rm{s}^{-1}$), $\xi_{100}=\xi/100$ ergs 
$\rm{cm}^{-2} \rm{s}^{-1}$) and $n_{e}$ is the electron density. 
For a volume filling factor of unity, the emitting material has a thickness 
$\Delta R \sim 1.2 N_{H}/n_{e}$ where $n_{e} \sim 1.2 n_{H}$ and $n_{H}$ is 
the density of Hydrogen nuclei (assuming He/H $\sim 0.1$ and He contributes 
two electrons). Therefore, if we assume that $\Delta R <R$, then 
$n_{e}>1.7 N_{21}^{2} \xi_{100}/L_{44} \rm{cm}^{-3}$ where $N_{21}=N_{H}/10^{21} \rm{cm}^{-2}$. 
The best--fit values of $N_{H}$ and $\xi$ from the XSTAR emission model fits 
then
allow us to establish a lower limit on the electron density in 
the emitting material.
It is important to note however, that the emitting material is
not necessarily the same as the absorbing material and therefore
may not have the same physical parameters or location as the
outflow responsible for absorption.

The three AGN spectra with the most prominent narrow emission lines 
were NGC~4051, NGC~5548 and NGC~3783. In the spectrum of NGC~4051, 
the \oxysevenf and \neninef forbidden lines 
were the most prominent, with additional emission due to \oxla and \nelya 
and weak \oxyseveni emission. There was no apparent \oxysevenr 
emission, indicating that the emitting material in this source is 
photoionized rather than collisionally ionized. Of course it is also 
possible that \oxysevenr emission may have been re-absorbed by \oxysevenr 
line absorption, which depends on the relative velocity shifts of the 
emitting and absorbing gases as well as the blending effects of the 
instrument resolution and the spectral binning. The forbidden lines 
were marginally 
blueshifted from systemic velocity and the other (less prominent) 
emission lines were consistent with emission at systemic velocity. The 
weighted mean velocity offset of the emission lines in this source 
($-160 \pm 75$ km $\rm{s}^{-1}$) was 
apparently blueshifted. 

NGC~5548 also exhibited strong \oxysevenf and 
\neninef forbidden emission lines, with relatively strong \oxla and 
\oxyseveni emission. \oxysevenr line emission was weak in this source, 
indicating that the emitter is predominantly photoionized rather 
than collisionally ionized. The He-like Ne and O triplet emission 
lines in NGC~5448 were 
marginally blueshifted (\oxysevenf lay at an offset velocity of 
$-265^{+160}_{-215}$ km $\rm{s}^{-1}$), but the 
\oxla emission was marginally redshifted. The weighted mean offset 
velocity of the emitting material ($-115 \pm 125$ km $\rm{s}^{-1}$) 
was consistent with emission at systemic velocity. NGC~3783 exhibited very 
strong \oxla, \neninef and \oxysevenr 
emission, the latter indicating that collisional ionization 
may be important in the emitting material \citet{b31}, although 
photoexcitation due to high levels of UV flux could also account for 
\oxysevenr emission \citet{b75}. Also
 present were emission features due to \oxysevenf and \oxyseveni. The He-like 
forbidden emission lines in NGC~3783 were marginally blueshifted, but the 
other emission lines were marginally redshifted. The weighted mean offset 
velocity of the emitting material was consistent with emission 
at systemic velocity. 

The spectra of Mkn~279, NGC4593, Mkn~509 and Mkn~766 each 
exhibited several emission features, most prominently \oxla and/or 
 \oxysevenf, but they did not exhibit $>2$ emission features at 
$>99\%$ significance. Of the ten AGN, 
we therefore only fit the spectra of NGC~3783, NGC~4051 and NGC~5548 with 
XSTAR models of emission. The best--fit results are given in 
Table~\ref{tab:em}. The requirement 
that $\Delta R/R <1$ yields lower limits of 
$n_{e}>3.7 \times 10^{6}, 2.3 \times 10^{2}, 
2.8 \times 10^{2} \rm{cm}^{-3}$ for NGC~4051, NGC~5548 and NGC~3783 
respectively.

\section{Properties of the Warm Absorbers in Type~I AGN}
\label{sec:wabs}
 
The results of the photoionization modelling of the AGN spectra are 
summarized in Table~\ref{tab:1wabs}. Figures~\ref{fig:3783}, 
~\ref{fig:ufspec1} and ~\ref{fig:ufspec2} show the best--fit photoionization 
models (from Table~\ref{tab:1wabs}) superimposed on the AGN spectra. Our 
uniform analysis of the AGN sample has yielded warm absorber parameters 
($N_{H}$, log $\xi$, velocity) that agree reasonably well with those in the 
literature (often from much more detailed analyses). 

Our results provide the first overview of high resolution soft 
X--ray grating data from \chandra observations of type~I AGN. In summary, 
we found that nine of 
the fifteen AGN in the sample have intrinsic continua that are more complex
than a simple power law in the 0.5--5.0 keV band (modelled here as a broken 
power law). Of the remaining six AGN spectra, the continua of three 
(F9, 3C 120 and Mkn~279) 
are well described by simple power law models and the continua of the three 
remaining AGN (NGC~3227, NGC~3516 and NGC~7314) are well described by 
heavily absorbed power law models. Note that previous, higher S/N data of the
 heavily absorbed AGN (NGC~3227 \& NGC~7314), evidence for photoionized 
absorption has in some cases been detected (see e.g. 
\citet{b93};\citet{b13}). The nine AGN spectra with complex intrinsic 
continua exhibit signatures of an ionized absorber as does the spectrum of 
NGC~3516, in spite of being heavily absorbed. The absorbers appear 
to be photoionized and outflowing, with velocities in the range 
$\sim 0-2000$ km $\rm{s}^{-1}$ (similar to the speed of the Solar wind). 
The column density of the warm absorbing gas is 
$\sim 10^{20-23} \rm{cm}^{-2}$. Nine of the ten AGN exhibiting 
warm absorption are best--fit by multiple 
ionization components and three of the ten AGN \emph{require} multiple 
kinematic components. The warm absorbing gas in our AGN sample has a 
wide range of ionization states, spanning roughly four orders of magnitude 
($\xi \sim 10^{0}-10^{4}$) and up to three orders of magnitude in the same 
source. Of the ten AGN spectra that exhibit warm absorption, a simple fit 
to the 0.5-0.7 keV and 2.5-5.0 
keV energy regions reveals that the intrinsic continuum of two AGN 
(NGC~4051 and NCG~5548) exhibits a statistically significant upturn in the 
intrinsic soft X--ray continuum relative to a hard X--ray power law. 
Our simple test shows that \mcg does not require a soft excess. 
The remaining AGN that exhibit warm absorption are either heavily absorbed 
in soft X-rays or exhibit spectral complexity that cannot be accounted 
for by such a naive test of the data. Nine of the ten AGN that exhibit warm 
absorption also exhibit Fe UTAs, indicative of an absorber component that 
could (depending on absorber density and geometry) carry most of the mass 
in outflow, as we show below.

\begin{figure}
\includegraphics[height=4in,width=3in]{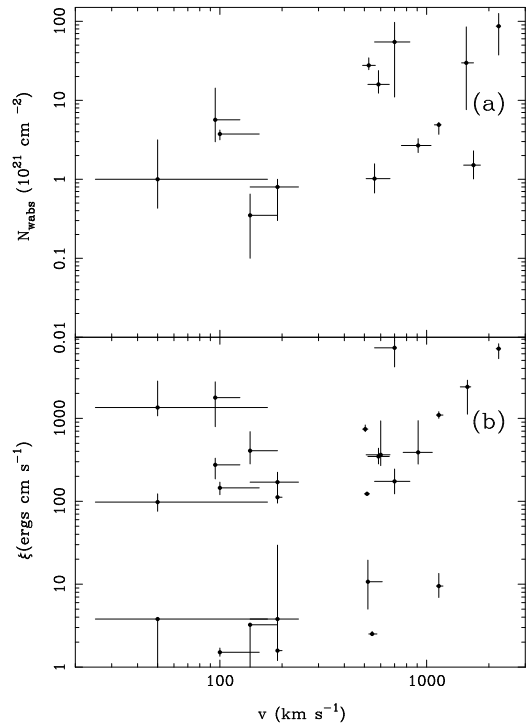}
\caption{Velocity of Warm Absorber Components (a) vs Column Density ($N_{H}$)
  and (b) vs $\xi$ of Warm Absorber Component for the AGN in our sample 
(from Table~\ref{tab:1wabs}). In panel (a), we plot the total column 
density of all the warm absorber components around a particular velocity. Thus, e.g. the 
three low velocity components from NGC 3783 in Table~\ref{tab:1wabs} become a
 single velocity component between -490 and -565 km $\rm{s}^{-1}$ in panel 
(a). In panel (b), we plot 
\emph{all} the individual warm absorber 
components to illustrate the fact that the ionization parameter can vary 
by up to three orders of magnitude in the same source. In panel (b), 
the low ionization components ($\xi <10.$) correspond to the Fe UTAs.
\label{fig:vwabs}}
\end{figure}

Fig.~\ref{fig:vwabs} compares the warm absorber component parameters from 
the different AGN (see Tables~\ref{tab:1wabs} \& \ref{tab:em}). 
Fig.~\ref{fig:vwabs}(a) shows the relationship between warm absorber outflow 
velocity and the corresponding warm absorber column density. 
Clearly the warm absorber components in all ten AGN are 
outflowing and with velocities that span around two orders of magnitude. 
There appears to be a gap in the outflow velocities
in our sample between $\sim 300-500$ km $\rm{s}^{-1}$, the origin of
which is not clear. The outflow components with velocities below this
gap tend to be associated with lower column densities than those with
with velocities above the gap. 
Fig.~\ref{fig:vwabs}(b) shows that the ionization parameter of the warm 
absorber is independent 
of velocity and can vary by up to three orders of magnitude in
the same source,
suggesting that the outflowing
absorber components
consist of gas lying at different points on the heating/cooling 
curve (e.g. \citet{b19}). Furthermore Fig.~\ref{fig:vwabs}(b) also indicates 
that there is a separate population of low ionization state 
absorber components at $\xi<10$. These low ionization parameter absorber 
components are responsible for the Fe UTAs in the AGN spectra. One caveat is 
that there is no bias in principle in the X-ray band for absorbers with 
ionization states $10<\xi<100$, however \chandra \hetg is less sensitive than 
other instruments to this band.

The warm absorber outflow apparent in ten of the fifteen AGN in our 
sample must clearly carry mass away from the central SBH. The rate of mass 
loss from an AGN outflow is given by 

\begin{equation}
\dot{M}_{\rm outflow}= \left(\frac{\Delta \Omega}{4\pi}\right) C y m_{p} n_{e} v 4 \pi r^{2}
\end{equation}

where $\Delta \Omega$ is the solid angle subtended
by the absorber at the ionizing source, 
$C$ is the volume filling factor, $y$ is the mean atomic mass per
Hydrogen atom ($y \sim 1.3$ for solar abundances),
 $m_{p}$ is the proton mass, $n_{e}$ is 
the electron density of the absorber, $r$ is the distance of the absorber from 
the radiation source,
and $v$ is the outflow velocity. 
We believe that the filling factor, $C$, cannot be reliably 
constrained by current observations of AGN (see e.g. discussion in 
\S\ref{sec:compare}). 
Blustin \etal (2005) estimated $C$ for a sample of sources
based on various assumptions, but those same assumptions
led Blustin \etal (2005) to derive a maximum distances between
the absorber and radiation source that was {\it less} than the
minimum distance for no less than five sources.
On the 
other hand, it is possible to argue that the covering factor 
($\Delta \Omega/ 4\pi$) is approximately
 equal to the fraction of type~I AGN in which a warm absorber is detected, 
or $\sim 0.5-0.7$. However, this argument breaks down if AGN
winds are accelerating and/or ``bend'', rather than constant velocity,
non-accelerating outflows.
Even so, this still leaves $C$ (and therefore the mass outflow
 rate) unconstrained. From the definition of the
ionization parameter, $r^{2}=L_{ion}/n_{e}\xi$, therefore

\begin{equation} 
\dot{M}_{\rm outflow}= 16.6 xy \left(\frac{\Delta \Omega}{4 \pi}\right) 
\frac{L_{ion,44}} {\xi_{100}} v_{500} C M_{\odot} yr^{-1}
\label{eq:out}
\end{equation} 

where $L_{ion,44}=L_{ion}/10^{44}$, $\xi_{100}=\xi/100$ and $v_{500}=v/500$ 
where $v$ is the outflow velocity, and $x$
is the number of Hydrogen atoms per free electron.
With $x=9/11$
 and $y=1.3$, the product $xy$ is approximately unity, or $\sim 1.06$.
Equation~\ref{eq:out}
suggests that all else being equal, warm absorbing gas with a low value of 
ionization parameter should account for the largest mass outflows from AGN. 
From Table~\ref{tab:1wabs} we find mass outflow rates in the range 
$\sim 10^{-1}-10^{3} \times (\Delta \Omega/4\pi) C M_{\odot} \rm{yr}^{-1}$. 
Therefore, even for 
small filling factors, the outflow rate can be comparable to the expected 
accretion rate onto the central supermassive black hole. The accretion rate 
onto the black hole can is $\dot{M}_{\rm accretion} = L_{\rm bol} / 
\eta c^{2}$, where 
$L_{\rm bol}$ is the bolometric luminosity of the AGN and $\eta$ is the 
accretion 
efficiency of the black hole. If $L_{\rm bol}= 10 X_{10} L_{ion,44}$, where 
$X_{10}$ is a
 parameter of order unity, and 
$\eta = 0.1\eta_{0.1}$, then $\dot{M}_{\rm accretion} = L_{ion,44} X_{10}/ (5.7 
\times \eta_{0.1}) M_{\odot} /yr$ and

\begin{equation}
\frac{\dot{M_{\rm outflow}} }{\dot{M_{\rm accretion}} } \sim 94 
\left(\frac{\Delta \Omega}{4\pi}\right) \left(\frac{xy}{X_{10}}\right)
\left(\frac{v_{500} }{\xi_{100}}\right) \eta_{0.1} C  
\label{eq:only}
\end{equation}

\begin{table*}
 \begin{minipage}{110mm}
  \caption{Black hole masses for AGN with warm absorbers \label{tab:bh}. 
Column 2 lists the 0.5-2.0~keV luminosity for each source, corrected for 
absorption. References: ($1$) \citet{b58}. ($2$) \citet{b64}. ($3$) 
\citet{b66}. ($4$) \citet{b56}.
}
\begin{tabular}{@{}lrrrrr@{}}
\hline
 Source & 0.5-2.0 keV & $L_{ion}$ & $M_{BH}$ &$L_{ion}/L_{edd}$ & Reference\\
        & luminosity &  	  & 	     & 			& \\
        & ($10^{43}$erg/s)&($10^{44}$erg/s) & ($10^{6}M_{\odot}$) & &\\
\hline
NGC~3516 (low) & 0.16 & 0.15&43& 0.003 & 1\\
NGC~3516 (high)& 0.38 & 0.36& \ldots & 0.007 & 1 \\
NGC~3783       & 1.25 & 1.00&30 & 0.003& 1\\
NGC~4051       & 0.02 & 0.02&2.0 & 0.006 & 1\\
Mkn 766        & 0.66 & 0.47&0.6 & 0.572 & 2\\
NGC~4593       & 0.50 & 0.29&5.4 & 0.042& 1\\
MCG-6-30-15    & 0.32 & 0.16&4.5 & 0.027& 3\\
IC 4329A       & 0.76 & 3.40&7.5 &0.349 & 1\\
NGC~5548       & 6.98 & 8.33&96&0.067 & 1\\
Mkn 509        & 6.55 & 5.23&100 &0.040 & 1\\
Akn 564        & 6.54 & 4.69&1.2 &3.000 & 4\\
\hline
\end{tabular}
\end{minipage}
\end{table*}

The value of $C$ and the covering factor
of the absorber is critical to understanding the 
processes underpinning the warm absorption phenomenon. If $v_{500}, X_{10}, 
\eta_{0.1} \sim 1$ and if $(\Delta \Omega /4\pi) \sim 0.5$ typically, 
then the filling factor must be very small ($< 0.02$) for the mass outflow 
rate to be comparable to or less than the
accretion rate (a point also made by \citet{b39}).



\begin{figure}
\includegraphics[height=3in,width=3in]{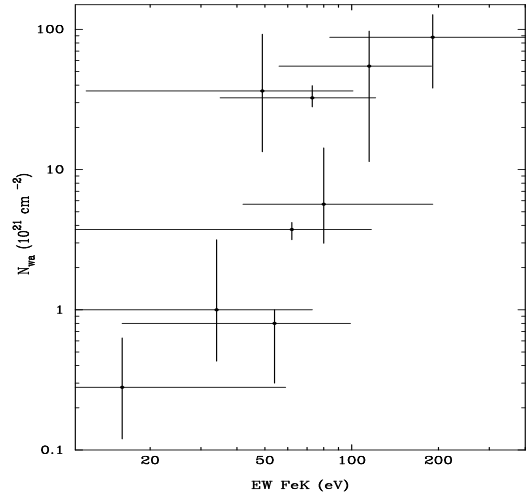}
\caption{EW of the Fe~K$\alpha$ line in each AGN versus the total warm 
absorber column density (from Table~\ref{tab:em}). Fe~K$\alpha$ measurements 
were taken from \citet{b43}. A weak correlation between total column 
density of absorber and the EW of the Fe~K$\alpha$ line is permitted, 
however Fe~K$\alpha$ line EWs can vary significantly in individual AGN, so 
a correlation is unlikely.
\label{fig:ewfek}}
\end{figure}

The warm absorber components found in Seyfert~1 galaxies have
column densities that are too small to account for the 
EWs of the narrow, core, Fe~K line emission. However, for our sample, we 
investigated whether the core Fe~K line emission EW correlate
with the column densities of the warm absorber components in case
of a secondary effect. For example, if the warm absorber material
is an outflow of material from the putative obscuring torus,
and if the core Fe~K line emission originates in the torus itself,
one might expect a correlation between the Fe~K line EW and the
warm absorber column density if thicker torii produce thicker
winds.
We also investigated whether the Fe~K$\alpha$ line in Type I AGN `knows' 
about the warm absorber by compared the total column density of the warm 
absorber in an AGN with the EW of the corresponding core Fe~K$\alpha$ 
line \citep{b43} (Figure~\ref{fig:ewfek}). 

For our sample, Figure~\ref{fig:ewfek} shows the EW measurements 
(from Yaqoob \& Padmanabhan 2004) versus warm absorber column density.
A weak correlation is 
permitted based on Figure~\ref{fig:ewfek}, so the Fe~K$\alpha$ line might 
`know' about a large warm absorber column density. In this case we calculate 
the Pearson correlation coefficient to be $0.88$ (for N=9), which is 
significant at the $\sim 99\%$ confidence level. However, 
we caution that the EWs of 
individual Fe~K$\alpha$ lines can vary by a factor of 2-3
as the continuum varies whilst the line intensity does not, and
this would 
destroy any hint of correlation in Figure~\ref{fig:ewfek},
which may therefore be spurious. In particular, we note that no warm absorber 
is detected in F9, yet it \emph{has} a prominent, strong Fe~K$\alpha$ 
line core \citep{b43}.

\subsection{Comparison with other AGN sample studies}
\label{sec:compare}

\citet{b73} collated from the literature the results of some of the 
X-ray spectral observations of 23 AGN using \chandra and \xmm and 
attempt to infer the general properties of Seyfert warm absorbers 
from these results. The methodology of the \citet{b73} study is quite 
different from the present work, 
However, the strengths of our study (uniform analysis and a single instrument) 
may also be weaknesses. Our selection criteria for warm absorption signatures 
may be too conservative, our uniform modelling of the continuum may ignore 
individual source peculiarities and our band-pass may miss e.g. very low 
ionization absorber components. A study such as the one carried out by 
\citet{b73} is a useful comparison for these reasons. \citet{b73} conclude 
that most warm absorbers are most likely to originate in outflows from 
the dusty torus, since they estimate that the minimum distance of the 
warm absorbers from the central radiation source mostly cluster around 
the distance of the torus. \citet{b73} conclude that the kinetic luminosity of 
these outflows is $<1\%$ of the AGN bolometric luminosity and the observed 
soft X-ray absorbing ionization phases fill $<10\%$ of the available volume.

Some of the conclusions reached by \citet{b73} are quite different from ours. 
We make no claim for the filling factors of the outflows or indeed, their 
origin. First, we believe there are currently insufficient data and 
constraints to reliably estimate the filling factor, preventing the 
derivation of absolute mass outflow rates. Second, we note that \citet{b73}
 calculate a minimum distance between the central radiation source and the 
warm absorber based only on the assumption that the outflow velocity exceeds 
the escape velocity. However, very recently \citet{b76}
 observed an X-ray absorbing outflow from the stellar-mass black hole binary 
GRO J1655-40 with a radial velocity \emph{far less} than the escape velocity 
at its location. Moreover, the assumptions of \citet{b73} lead
to a maximum distance of the warm absorber from the
radiation source that is {\it less} than the minimum distance 
for no less than five AGN.
Robust information on the distance between the warm absorber 
and the central source can only come from variability studies of the absorber. 
Third, \citet{b73} conclude that Seyfert warm absorbers 
are probably not telling us anything fundamental about the energetics or 
structure of the central engine. However, if for example, warm absorbers 
originate in a disk wind, the properties of the disk wind are 
likely to correlate with those of the disk (magnetic flux strength, disk 
temperature gradient, velocity of disk etc). Finally, some parts of the warm 
absorbing outflows from AGN discussed here and by \citet{b73} \emph{may 
not actually be observable} because they are fully ionized. Such fully 
ionized outflows may be the fastest component of the outflow since it may be 
closest to the source of ionizing radiation.

\section{Discussion}
\label{sec:discussion}

At present there are two distinct theories of the origin of the warm 
absorber. On the one hand, \citet{b11} proposes a unified scheme for 
quasars that includes a wind that 
rises vertically from a narrow range of radii on the inner accretion disk 
and is then bent outwards by a radial radiation force to produce a 
funnel--shaped thin shell outflow. This model specifically excludes the 
dusty molecular torus as a characteristic of AGN and proposes that the  
Fe~K$\alpha$ emission line is produced in the funnel. 
\citet{b11} predicts that 
the range of line widths will be similar to the broad absorption line 
`detatchment velocities' ($\sim 0-5000$ km $\rm{s}^{-1}$ vertically, so one 
might expect a narrower line width than this when looking down the funnel). 
Of the forty-eight absorption features used to constrain the warm absorber 
components in this study, thirty-seven have FWHM $<$ 2000 km 
$\rm{s}^{-1}$ and only four allow for FHWM $>3000$ km $\rm{s}^{-1}$. 
Statistically, a weak correlation is allowed between the EW of the FeK$\alpha$
 line and the total column density of the warm absorber. However, we note that
 in our sample, the source with the least evidence for \emph{any} warm 
absorption (F9) exhibits the largest FeK$\alpha$ EW, which is not consistent 
with expectations from the Elvis model.

On the other hand, 
\citet{b19} propose that the warm absorber originates in a photoionized 
evaporation from the inner edge of the putative obscuring torus 
believed to surround the AGN central engine. In this model, 
the warm absorber is a multi-temperature wind,
with the different outflow components having values
of $\Xi$ (pressure form of ionization parameter) that cover
a relatively narrow 
range, as a result of
the different phases of gas co-existing in pressure equilibrium.
From our photoionization model fits, we find that the temperatures
of the warm absorber components lie in the range $T \sim 10^{4-7} \ \rm K$
for the sample, yet $\log{\Xi}$ lies in the range $\sim 0-1$, with
the highest ionization components generally having the largest
values of $T$ and $\Xi$. Thus, our general constraints seem to agree quite 
well with the model predictions of \citet{b19}. 
An additional attraction of 
this model is that the interaction of the disk wind with the 
dusty molecular torus in this model can also naturally explain dusty 
warm absorption, which may be the cause of additional spectral 
complexity of some of the AGN in this sample (see \S\ref{sec:soft}).

One important point to note is that some of the 
warm absorber components discussed above may actually be due in part to 
hot local gas at cz$\sim$0. For example the low velocity warm 
absorber component in NGC~4051 at ($\sim -600$ km $\rm{s}^{-1}$) 
is kinematically very close to an absorption signature at cz$\sim$0 since 
NGC 4051 
is cosmologically redshifted by $cz=726 \pm 15$ km $\rm{s}^{-1}$ from 
z$\sim$0. 
Other AGN in this sample yield tantalizing hints of absorption due to hot 
local gas (see \citet{b52,b47} and references therein 
for further details). Of course, the absorption 
signatures of some of the warm absorbing outflows could be mimicked by hot gas
 at intermediate redshift. However, the expected column density of most 
filaments of warm/hot intergalactic medium \citep{b70} is far lower than that 
observed by 
us.

The soft X-ray spectra of type~I AGN observed with the high resolution
 \chandra and \xmm observatories prompt many intruiging questions. We believe 
that these questions can be answered only by the careful uniform analysis of a 
sample of AGN. Our uniform analysis has its limitations since we use spectra 
from only one instrument with a limited bandpass, and our methodology is 
sufficiently conservative that we may miss less signficant absorber 
signatures. Nevertheless, we have carried out a first uniform analysis on a 
small sample of AGN observed with \chandra \hetg and we have established 
reasonably well constrained parameter ranges for the warm absorbing 
outflows in type I AGN. Interestingly, we found that mass loss resulting 
from the warm absorber outflow can be high, 
comparable to or greater than the expected accretion rate onto the central 
supermassive black hole. Most of the outflowing mass could be carried by low 
ionization 
state outflows (depending on absorber density and geometry), which are 
best studied in the X-ray band via their Fe UTA spectral imprint. Low 
rates of mass outflow from AGN and higher velocity outflows may be 
associated with lower values of the AGN Eddington ratio, but the rate 
of mass outflow does not correlate with the mass of the central black 
hole. A weak contribution to the core of the narrow Fe~K$\alpha$ line from 
high column density warm absorbers seems unlikely. There is gap in the
outflow velocities of the warm absorber components in the range
$\sim 300-500$  km $\rm{s}^{-1}$ that is puzzling.
All these results so far provide 
tantalizing hints at global patterns of type I AGN behaviour and
merit further investigation using 
a larger sample of high spectral resolution AGN spectra.  

\section*{Acknowledgments}
BM gratefully acknowledge support from NSF grant AST0205990. TY acknowledges 
support from NASA through grant AR4-5009X issued by 
Chandra X--ray Observatory Center, operated by the SAO for and on behalf of 
NASA under contract NAS8-39073. We made use of the HEASARC on-line data 
archive services, supported by NASA/GSFC and also of the NASA/IPAC 
Extragalactic Database (NED), operated by the Jet Propulsion Laboratory, 
CalTech, under contract with NASA. Thanks to the \chandra instrument and 
operations teams for making the observations possible. Thanks to Tim Kallman 
for numerous useful discussions on XSTAR. Thanks to the anonymous referee for 
very detailed and useful comments that helped improve and shorten this paper. 
Thanks to Fabrizio 
Nicastro for discussing his results from an unpublished \xmm RGS observation 
of NGC~4051.

\bsp

\label{lastpage}

\end{document}